\documentclass[12pt,preprint]{aastex}

\shorttitle{SWIRE Galaxy Populations}
\shortauthors{Lonsdale et al.}

\received{2004 April 5}
\begin{document}

\title{First Insights into the SWIRE Galaxy Populations}

\author{Carol Lonsdale}
\affil{Infrared Processing \& Analyis Center, California Institute 
of Technology, 100-22, Pasadena, CA 91125, USA} 

\author{Maria del Carmen Polletta}
\affil{Center for Astrophysics \& Space Sciences, University of 
California, San Diego, La Jolla, CA 92093--0424, USA}

\author{Jason Surace, Dave Shupe, Fan Fang \& C. Kevin Xu}
\affil{Infrared Processing \& Analyis Center, California Institute 
of Technology, 100-22, Pasadena, CA 91125, USA}

\author{Harding E. Smith \& Brian Siana}
\affil{Center for Astrophysics \& Space Sciences, University of 
California, San Diego, La Jolla, CA 92093--0424, USA} 

\author{Michael Rowan-Robinson \& Tom Babbedge}
\affil{Astrophysics Group, Blackett Laboratory, Imperial College,
Prince Consort Road, London, SW7 2BW, UK} 

\author{Seb Oliver, Francesca Pozzi \& Payam Davoodi}
\affil{Astronomy Centre, CPES, University of Sussex, Falmer, 
Brighton BN1 9QJ, UK}

\author{Frazer Owen}
\affil{National Radio Astronomy Observatory, P.O. Box O, Socorro, 
NM 87801, USA} 

\author{Deborah Padgett, Dave Frayer, Tom Jarrett, Frank Masci, 
JoAnne O'Linger \& Tim Conrow}
\affil{Infrared Processing \& Analyis Center, California Institute 
of Technology, 100-22, Pasadena, CA 91125, USA}

\author{Duncan Farrah, Glenn Morrison \& Nick Gautier}
\affil{Jet Propulsion Laboratory, 264-767, 4800 Oak Grove Drive,
Pasadena, CA 91109, USA}

\author{Alberto Franceschini \& Stefano Berta}
\affil{Dipartimento di Astronomia, Universita di Padova, Vicolo 
Osservatorio 5, I-35122 Padua, Italy} 

\author{Ismael Perez-Fournon}
\affil{Instituto de Astrofisica de Canarias, 38200 La Laguna, 
Tenerife, Spain} 

\author{Herve Dole}
\affil{Institut d'Astrophysique Spatiale, Universite 
Paris Sud bat 121, F-91405 Orsay, France} 

\author{Gordon Stacey}
\affil{Department of Astronomy, Cornell University, 220 Space 
Science Building, Ithaca, NY 14853, USA} 

\author{Steve Serjeant}
\affil{Centre for Astronomy and Planetary Science, School of 
Physical Sciences, University of Kent at Canterbury, Canterbury, Kent CT2 7NZ} 

\author{Marguerite Pierre}
\affil{CEA/DSM/DAPNIA, Service d'Astrophysique, 91191 
Gif-sur-Yvette, France} 

\author{Matt Griffin}
\affil{Department of Physics and Astronomy, University of Wales 
Cardiff, 5 The Parade, Cardiff CF24 3YB, UK} 

\author{Rick Puetter}
\affil{Center for Astrophysics \& Space Sciences, University of 
California, San Diego, La Jolla, CA 92093--0424, USA}


\begin{abstract}
We characterize the SWIRE galaxy populations in the SWIRE validation field 
within the Lockman Hole, based on the 3.6-24$\mu$ Spitzer data
and deep U,$g^\prime$,$r^\prime$,$i^\prime$ optical imaging within an area 
$\sim$1/3 sq. deg for $\sim$16,000 Spitzer-SWIRE sources.  
The entire SWIRE survey will discover over 2.3 million galaxies at 
3.6$\mu$m and almost 350,000 at 24$\mu$m; $\sim$ 70,000 of these will be 
5-band 3.6-24$\mu$ detections.
The colors cover a broad range, generally well represented by 
redshifted spectral energy distributions of known galaxy populations, however 
significant samples of unusually blue objects in the [3.6-4.5]$\mu$m color
are found, as well as many objects very red in the 3.6-24$\mu$m mid-IR.  
Nine of these are investigated and are interpreted as star-forming systems,
starbursts and AGN from z=0.37 to 2.8, with luminosities from 
L$_{IR}$=10$^{10.3}$ to 10$^{13.7}$ L$_{\odot}$. 

\end{abstract}


\section{Introduction and Observations}

The Spitzer Wide-area InfraRed Extragalactic Legacy Survey, SWIRE
(Lonsdale et al. 2003), will map the evolution of spheroids, disks, 
starbursts and AGN to z$>$2, within volumes large enough 
to sample the largest important size scales.   We present initial 
results from deep optical (U,$g^\prime$,$r^\prime$,$i^\prime$) and 
Spitzer-SWIRE (3.6$\mu$m-24$\mu$m) imaging of 0.3 sq. deg. in the SWIRE 
Survey validation field (VF) in the Lockman Hole,
a field selected to have extremely low cirrus emission, and a lack of 
bright radio sources.  Deep K-band and VLA 20 cm imaging also exist, and 
this field will be 
imaged with Chandra/ACIS-I to 70ks depth in 2004 August. 
The full SWIRE survey will image $\sim$49 sq. deg. in all IRAC and MIPS bands 
in 6 fields, Area has been reduced from the strategy described 
by Lonsdale et al. (2003) in order to maintain
two high quality coverages of each field with the MIPS 70$\mu$m array 
(see {\tt http://www.ipac.caltech.edu/SWIRE} for details). 
The SWIRE validation field was imaged by Spitzer in December 2003 following
the strategy 
described in Lonsdale et al. 
(2003), therefore it has shallower MIPS depth than the main SWIRE survey.  
The full SWIRE Lockman field was imaged with the new strategy 
in April/May 2004.   

The SWIRE VF is centered at $10^h46^m, +59^d01^m$.
The observations were executed on 2003 Dec 05 \& Dec 09.
The Spitzer PROGID for these data is 142 and the datasets are
identified as IRAC:
\dataset[ads/sa.spitzer\#0007770880]{(AOR key 7770880)}
\dataset[ads/sa.spitzer\#0007771136]{(AOR key 7771136)}
MIPS:
\dataset[ads/sa.spitzer\#0007770368]{(AOR key 7770368)}
\dataset[ads/sa.spitzer\#0007770624]{(AOR key 7770624)} .
Data processing began with the Spizer Basic Calibrated Data products, which
are individual Spitzer images corrected for bias offsets and
pixel-to-pixel gain variations (flat-fielding), and 
flux-calibrated in surface brightness units of MJy/sr.  Additional
individual IRAC image processing corrected latent images and electronic
offset effects.  For MIPS, scan-mirror-dependent flats were derived from
the data and applied to the BCD images.  The individual images, which have
measurable spatial distortions, were reprojected onto a single common
projection system on the sky, and then coadded through averaging with outlier
rejection to remove cosmic ray and other transient artifacts.  
A 3-color 3.6, 4.5, 24$\mu$m false color
image of part of the field is shown in Plate 1.   

Fluxes were extracted in 5.8\arcsec\ apertures for IRAC 
($\sim$2-3$\times$ the FWHM beam) and 12\arcsec\ for MIPS 24$\mu$m,
using SExtractor.  
Very few ($<$5\%) of the detected objects are extended relative to the 
large Spitzer beams ($>$2\arcsec\ at the shortest wavelength), and even fewer on
scales comparable to the extraction apertures.  Aperture corrections have
been derived from stellar sources in the mosaicked data by the instrument
teams, and these have been applied to correct
to total fluxes.   The IRAC flux calibration is believed to be 
correct within 3\%, and the 24$\mu$m calibration to 10\%.  There is an 
additional scatter resulting from color dependencies in the flat 
field that add roughly a $<$10\% 
random error to the fluxes for all IRAC data.  The calibration was 
confirmed for IRAC
by comparison to 2MASS, a robust extrapolation from the 2MASS K-band
since the IRAC bands lie on the Rayleigh-Jeans tail of the stellar SED.
For 10 stars in our field with 2MASS
magnitudes, we extrapolated to a 24 micron flux density using
Kurucz-Lejeune atmospheric models for MK class I, III and V 
implemented in the SSC's Stellar Performance Estimation Tool, assuming
G5 spectral type, confirming the calibration to better than 10\%.  The 
resulting catalogs were examined by eye and remaining spurious sources (from 
radiation, scattered light, etc.) were removed by hand.
Details of the data processing are given in Surace et al (2004) and Shupe 
et al. (2004).

The optical $g^\prime$, $r^\prime$ and $i^\prime$ data were taken in Feb, 
2002 and the U data in Jan, 2004, using the MOSAIC camera on the Mayall 4m 
telescope
at Kitt Peak National Observatory (Siana et al. 2004, in preparation).  
The data were processed with the 
Cambridge Astonomical Survey Unit's reduction pipeline following the
procedures described by Babbedge et al. (2004).  Fluxes were extracted within
2.06\arcsec\ apertures and corrected for the aperture using profiles measured
on bright stars.    An analysis comparing the 2.06\arcsec\
aperture-corrected magnitudes with total ``Kron'' magnitudes in the $r^\prime$
band indicated that brighter than $\sim$21.5 mag. (Vega) significant numbers of galaxies 
have fluxes underestimated by the aperture photometry, therefore our analysis 
is limited to galaxies fainter than this limit.  Colors for the source samples 
were constructed from the aperture-corrected magnitudes.
   
The final depths for the Spitzer sample, at $\sim$5$\times$ the noise level, 
are given in Table 1. These depths are consistent with the 90\% completeness 
limits as determined from the deviation of the observed number counts
from a power law, and from simulated extractions
of artificial sources injected into the data.
The median achieved depths for extended objects (galaxies) in the optical 
bands, at 
the $\sim$90\% completeness levels for source extraction, derived in a similar
fashion to the Spitzer data, are U=24.9,  $g^\prime$=25.7,  $r^\prime$=25.0,  
$i^\prime$=24.0.

\clearpage
\begin{deluxetable}{lccccccccc}
\tabletypesize \footnotesize
\tablewidth{0pt} 
\tablecaption{Spitzer Sensitivities and Detection Statistics}
\tablehead{
\colhead{${\lambda}_c$ ($\mu$m)} & \colhead{3.6} & \colhead{4.5} & \colhead{5.8} & \colhead{8} & \colhead{24} & 
 \colhead{3.6,4.5} & \colhead{All IRAC} & \colhead{All IRAC,24} \\
}
\startdata
5$\sigma$ limit, validation field ($\mu$Jy) & 3.7 & 5.3 & 48 & 37.7 & 150 & \nodata & \nodata & \nodata \\
All detections, 0.3 sq. deg., VF & 16,075 & 12,675 & 1,536 & 1,657 & 1,290 & 11,765 & 950 & 433 \\
Galaxies (stars removed statistically) & 14,630 & 11,403 & 1,091 & 1,247 & 1,283 & 10,706 & 760 & 430 \\    
5$\sigma$ limit, full survey ($\mu$Jy) & 3.7 & 5.3 & 48 & 37.7 & 106 & \nodata & \nodata & \nodata \\
Predicted galaxies, 49 sq. deg. ($\times$10$^6$) & 2.39 & 1.86 & 0.18 & 0.20 & 0.35 & 1.75 & 0.12 & 0.07 \\
Model, 49 sq. deg., Xu et al S3$+$E2 & 6.36 & 4.76 & 0.24 & 0.33 & 0.72 & 4.75 & 0.21 & 0.17 \\
Projected/Model & 0.38 & 0.39 & 0.75 & 0.61 & 0.49 & 0.37 & 0.57 & 0.41 \\
\enddata
\end{deluxetable}
\clearpage

Stars were removed statistically from the Spitzer catalog using predicted 
counts based on a near-infrared Galactic stellar distribution model 
by Jarrett et al. (1994), based on the classic optical model of
Bahcall \& Soneira (1980).  The model was verified using
optical stellarity measures for objects associated with IRAC sources, and 
by matching star counts within a deep 2MASS catalog which reaches 
$\sim$2 mag. fainter than the all-sky 2MASS survey (Beichman et al. 2003).
Stars outnumber galaxies at F(3.6)${\ga}$150$\mu$Jy at this latitude 
(Surace et al. 2004) and decrease rapidly in number relative to galaxies 
below that flux density.

\section{Results and Discussion}

The detection statistics in Table 1 indicate that over the full SWIRE 
survey area of $\sim$49 sq. deg., we will detect $\sim$2.4 million galaxies at 
3.6$\mu$m and $\sim$120,000 in all 4 IRAC bands.  At 24$\mu$m the detection
rate for the full survey will be better than for the VF discussed here
because MIPS integration time has been doubled, therefore we estimate nearly
350,000 galaxies detected in this band, and about 70,000 of these in all
4 IRAC bands as well.   At optical wavelengths we detect the following numbers
of sources at U$g^{\prime}r^{\prime}i^{\prime}$ respectively: 27,911, 42,817, 
39,308, 30,230, stars plus galaxies.   17,894 are detected in all 4 optical
bands, and 8,626 in the combination $r^\prime$, 3.6, 4.5$\mu$m.  325 are detected
in all 9 optical+IR bands. 

We have compared the IR detection statistics to predictions from the  models 
of Xu et al.  (2003), model S3+E2, in Table 1.  Model S3 includes dusty 
objects: spirals, 
starbursts and AGN; and E2 contains passively evolving stellar systems, ie. 
spheroids.  The Xu et al. model overpredicts SWIRE IR galaxy 
numbers by a factor of $\sim$2 in the IR bands.  The number counts results 
will be addressed by Surace et al. (2004) and Shupe et al. (2004).  

Figures 1 and 2 present color-color plots which characterize the sample in
$g^\prime$ to 24$\mu$m color-color space.  Only sources detected in all
four bands shown in each figure are plotted; no limits are shown 
for clarity.  The figures show
SED-redshift tracks of several galaxies with a broad range of intrinsic
colors.  These SEDs cover the range of colors exhibited by known objects
throughout the U-24$\mu$m wavelength range and 0$<$z$<$2.  We do not expect
many sources in the region of the figures occupied by rare objects like Arp
220 at low redshift, because our volume coverage at low redshift space is
small.  A complete analysis of SWIRE galaxy colors relative to model
predictions and SED-tracks is beyond the scope of this paper, requiring
thorough analysis of selection effects, photometric redshifts and
k-corrections.  Here we note a few basic results.

There is a very broad distribution of colors in these figures.  Galaxies
with little on-going star formation will be relatively blue in the mid-IR
due to lack of dust emission, and also quite red in [$g^\prime-r^\prime$]
due to domination by late-type stars, and thus found towards the lower
right of Figure 1, near the elliptical SED track (red curve).  Indeed,
there is a concentration of systems near this region.  Moreover the
3.6$\mu$m-brightest systems in the sample (blue symbols) preferentially
inhabit this region, indicating that these may be relatively nearby
early-type systems.  The stellar tracks also cross this region of Figure 1:
using the stellar model described above we predict a maximum 0.13 star
fraction in the 10$<$F(3.6)$<$150 $\mu$Jy flux range, and 0.09 for
7.3$<$F(3.6)$<$10 $\mu$Jy, focused strongly within $\pm$0.2 magnitudes of
the stellar sequences.  In Figure 2 the elliptical SED track lies off the
figure to the bottom, due to lack of 24$\mu$m emission; objects in this
lower-right area are likely to be early-type spirals or unusually
dusty spheroids.

Dusty systems will be more strongly detected at the longer wavelengths, and
therefore redder in the Spitzer [3.6$-$4.5] and [3.6$-$24] colors.  There
is a trend in both figures that these systems tend also to be the bluest in
[$g^\prime-r^\prime$], inhabiting the upper left of both figures.  This is
expected for systems which have both  young complexes of
dust-enshrouded star formation dominating the mid-IR plus either (a)
hot blue young stars visible in lower optical depth regions 
at optical wavelengths, or (b) a blue type 1 AGN
visible in the optical, such as Mrk231 which tracks into this area at z$>$2
in Figure 2.  It is notable that the most
extreme systems (those toward the upper left of the figures), tend to be the
fainter galaxies in the sample at 3.6$\mu$m (red symbols).  This could be 
interpreted as due to either preferentially more distant systems or 
lower luminosity systems, however the complex selection and k-correction
effects would need to be understood in order to investigate this further.

\clearpage
\begin{figure}
\plotone{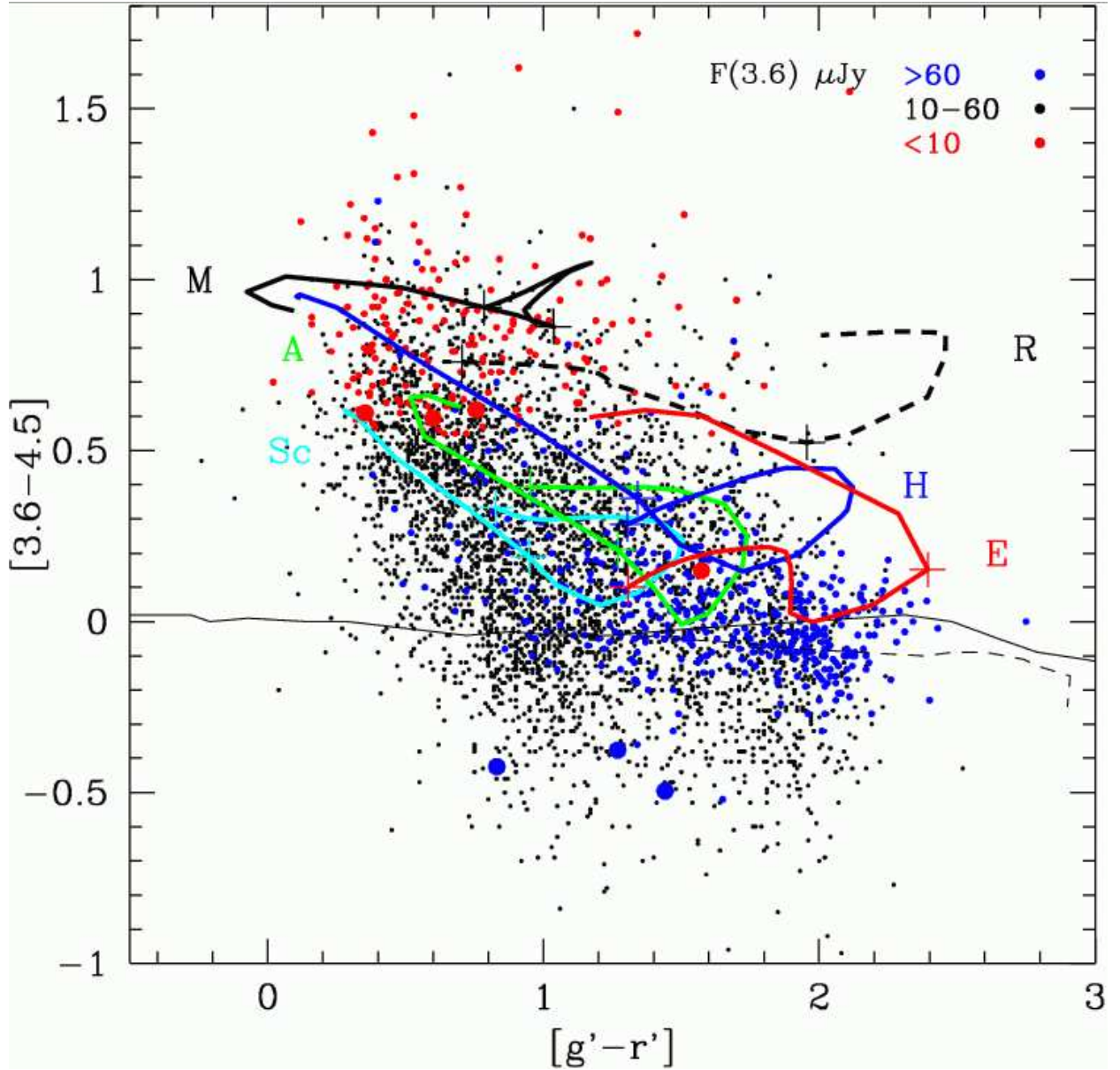}
\caption{Color-color distributions for 4,395 sources detected at 
$g^\prime$, $r^\prime$, 3.6 and 4.5$\mu$m with S/N$>$10 and 
$r^\prime{\ge}$21.5.  Sources with upper limits are not shown for 
clarity.   Stellar tracks: light black lines:  
main sequence solid; giants dashed;  Galaxy SEDs, 0.1$<$z$<$2, with crosses 
at z=0.1 and z=1:  elliptical (red; E); Sc spiral (cyan; Sc); Arp 220 
(green; A); the well-known ULIRG \& QSO Mrk 231 (black; M); an ERO, 
HR10 (blue; H); and a red QSO discovered in the FIRST survey 
(FIRST J013435.7-093102; optical/NIR spectrum from Gregg {\it et al.} (2002); 
at longer wavelengths the IR spectrum of PG1351+236, which has a very similar 
optical/NIR spectrum, was added by Polletta et al. 2004 (in preparation)),
(dashed black: R).  Units are Vega magnitudes, 
using zero points of 277.5, 179.5, 116.6, 63.1 \& 7.3Jy at 3.6, 4.5, 5.8, 
8 \& 24$\mu$m respectively.  Large solid dots indicate objects 
illustrated in Fig. 3 which are detected in each band depicted in this
figure: blue symbols for objects with [3.6$-$4.5]$<-$0.3 
(F(3.6)/F(4.5)$>$2.0) and red symbols for objects with [3.6-24]$>$7.5 mag 
(F(3.6)/F(24)$<$0.04).
}
\end{figure}
\clearpage
\begin{figure}
\plotone{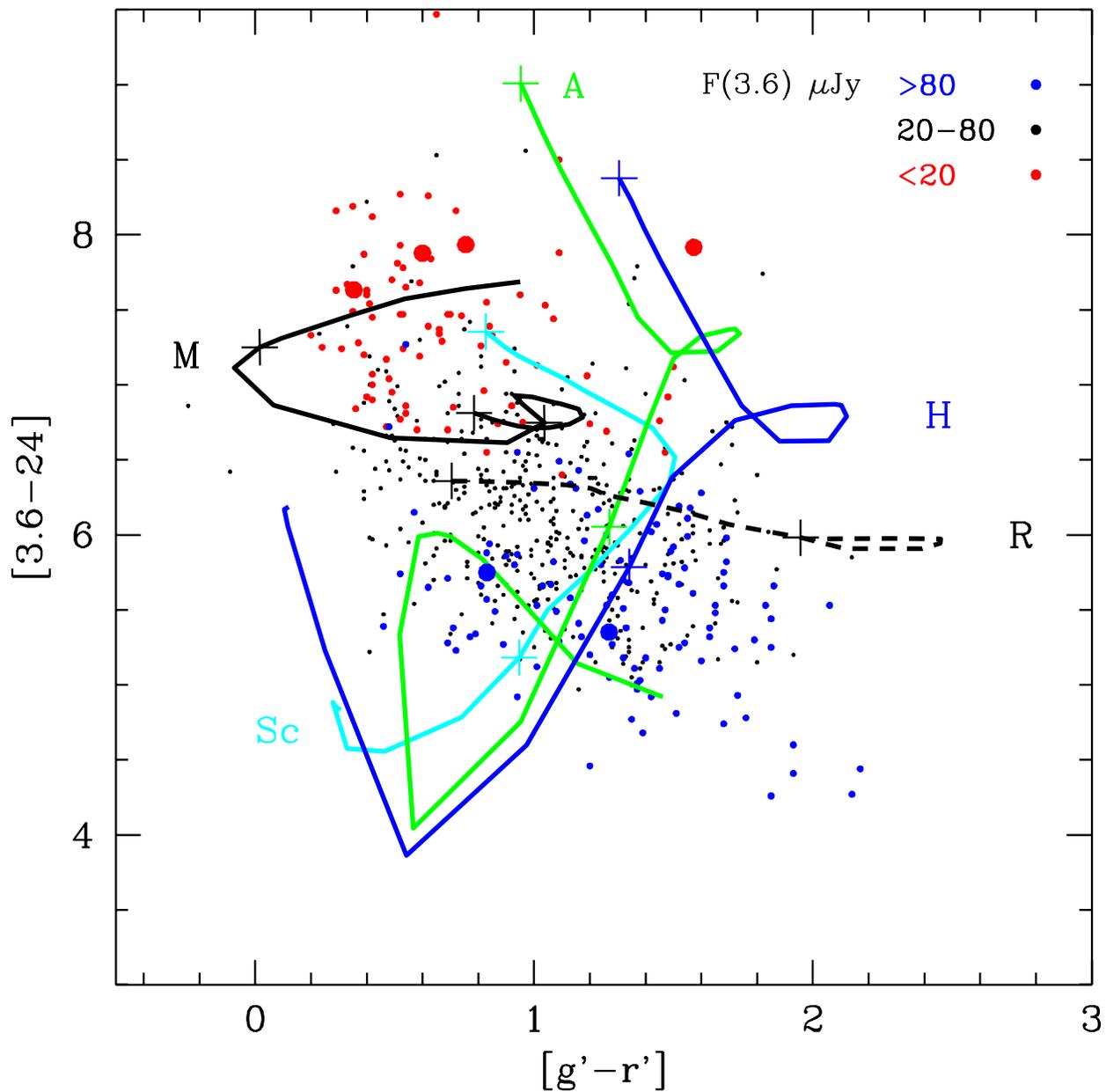}
\caption{Color-color distributions for 588 sources detected at 
$g^\prime$, $r^\prime$, 3.6 and 24$\mu$m with S/N$>$5 and 
$r^\prime{\ge}$21.5.  Tracks and symbols are as for Figure 1, except 
Mrk231 has been extended to z$=$3 with a additional marker plotted at z$=$2.
}
\end{figure}
\clearpage

Some areas of the color-color diagrams are not well covered by the SED
tracks.  Of particular note are some unusually blue objects in [3.6$-$4.5],
and mid-IR red sources at the upper left of Figure 2. 
Many additional extreme-colored objects with upper limits in one or more color
are not shown in these figures.
As an illustration of some of the most unusual objects populating 
the SWIRE sample, we have
investigated a number of these sources with red [3.6$-$24] colors 
and unusually blue [3.6$-$4.5] colors, 
using the photometric redshift code Hyper-z (Bolzonella et al. 2000) 
to fit SEDs with a wide range of templates, redshifts and A$_V$.   
We used our own library (Polletta et al. 2004),
the GRASIL library (Silva et al. 1998) and the Rowan-Robinson (2003) 
library.   The Polletta et al. (2004) library contains $\sim$40 
1000$\AA$ $-$ 20$cm$ templates 
for ellipticals, spirals, irregulars, starbursts, ULIRGs and AGN,
derived from observed SEDs, including mid-IR ISO spectra and 
models following Berta et al. (2004).  A more complete characterization
and photometric redshift
analysis of a larger SWIRE galaxy sample is in preparation 
(Rowan-Robinson et al. 2004).

Investigating first the blue sources, we selected 603 sources 
with [3.6$-$4.5]$<-$0.3, significantly bluer than normal galaxies and stars, 
with S/N$\geq$10 in both bands.   Fluxes were remeasured by hand for 193 of 
these objects with detections in a sufficient 
number of bands for SED analysis.  We used the IPAC-Skyview software to
set background levels interactively, thus avoiding confusion
with nearby sources and background contamination.
67 sources were found to have valid colors.  In about 
8\% of the remaining cases the 
automated source extractor measured a different region of a close
or confused pair of sources, or of an extended source, in the two bands. 
32\% of the sources marginally 
miss the color cut on careful color re-evalution, 8\% 
are cosmic rays or bad pixels, 36\% have anomalous 
3.6$\mu$m fluxes caused by local background or other effects due to 
bright stars, and 16\% have anomalous fluxes at either 3.6 or 
4.5$\mu$m with no 
obvious explanation.  The last 3 categories
represent a 76/12784=0.6\% anomaly rate amongst the entire S/N$>$10 3.6$\mu$m 
catalog, and a 40\% anomaly rate amongst the [3.6$-$4.5]$<$-0.3 sample.
Since anomalies are preferentially expected amongst odd-colored sources, the 
high anomaly rate amongst the unusually blue sources is not unanticipated.

The 67 valid [3.6$-$4.5]$<-$0.3 objects exhibit a wide range of 
optical to mid-IR colors.  7 are relatively blue and pointlike throughout the 
optical and mid-IR, and are probably stars. None of the galaxy libraries 
contain any templates as blue as the remaining [3.6$-$4.5]$<-$0.3 sources at 
any redshift.  Two objects in the literature have colors possibly
as blue at 3-5$\mu$m: the peculiar 
QSO [HB89]0049-29 at z=0.308 (Andreani et al. 2003), and the Seyfert 2 
ULIRG IRAS 00198-7826 at z=0.073 (Farrah et al. 2003).  [HB89]0049-29
peaks strongly in the NIR, however it is very red from there into the
optical, unlike any of our sources.  IRAS 00198-7826 is not observed
at these wavelengths, but is predicted to be as blue as our sources 
(Farrah et al. 2003), which is explained by it being a $\gtrsim60$Myr starburst
where much of the gas and dust has been blown away by 
supernovae.   This model for IRAS 00198-7826 can produce colors similar to
those of our blue objects in the [3.6$-$4.5] color, but it is too blue into
the optical to match any of our sources at any redshift.

Another possible explanation for the blue [3.6$-$4.5] colors is a strong
3.3$\mu$m PAH feature in the 3.5$\mu$m band at low redshift ($<$0.1), however
it would have to be considerably higher equivalent width than any such feature 
found in any of our templates.   Also possible is a strong 2.35$\mu$m 
CO bandhead absorption moving into the 4.5$\mu$m filter at redshift $\sim$0.7,
requiring a young stellar
population of red supergiants which is not diluted strongly by an older stellar
population with a weaker absorption (Rhoads 1997).  This might perhaps 
indicate a 
dominant $\sim$10$^7$ year old starburst in a fairly low mass galaxy.
Alternatively such a high equivalent width may indicate low metallicity.
 
In Figure 3 and Table 2 we present representative best fits for 5 blue sources 
(lower 5 SEDs; first 5 table entries).  The 24$\mu$m data points were down-weighted 
in these fits
so that they would not throw off the fit in the 3-5$\mu$m region
which we are primarily concerned with here; MIR SEDs can have a wide range of 
shapes
depending on details of geometry and astrophysics, which cannot be 
encapsulated in small libraries.  The fitted 
redshifts range from 0.68 to 0.94, and the corresponding infrared luminosities
range from log L$_{3-1000{\mu}m}$ = 10.3 to 11.3 L$_{\odot}$\footnote{H$_0$=71 
km/s/Mpc; ${\Omega}_m$=0.27; ${\Omega}_{\Lambda}=0.73$}; these are star 
forming galaxies and starbursts at moderate redshifts with moderate 
luminosities.   The blue [3.6$-$4.5] region of the SED is only
approximately fit, as anticipated, with deviations 0.8 to 
2.2$\sigma$ high for the 3.6$\mu$m points, and 4.3 to 7.9$\sigma$ 
low for the 4.5$\mu$m data points (combined deviations of the [3.6$-$4.5]
color from the template are given in the last column of Table 2).  
We present these fits as illustrative and not unique; 
fits at substantially different redshifts are possible
with different combinations of templates and A$_V$ values.  
If this phenomenon is confirmed as a real
and substantial population with unusually blue 3-5$\mu$m SEDs, ideal fits
will require modified template modeling outside the current 
libraries.  We note that the fitted redshifts for all of these objects
are consistent with the hypothesis of a dominant population of red supergiants
with strong CO absorption at 2.35$\mu$m redshifted into the 4.5$\mu$m
band.  It will be most interesting to discover whether Spitzer finds similarly 
blue colors in any regions within nearby galaxies, where the stellar 
populations and interstellar medium can be investigated in some detail.

\clearpage
\begin{figure}
\plotone{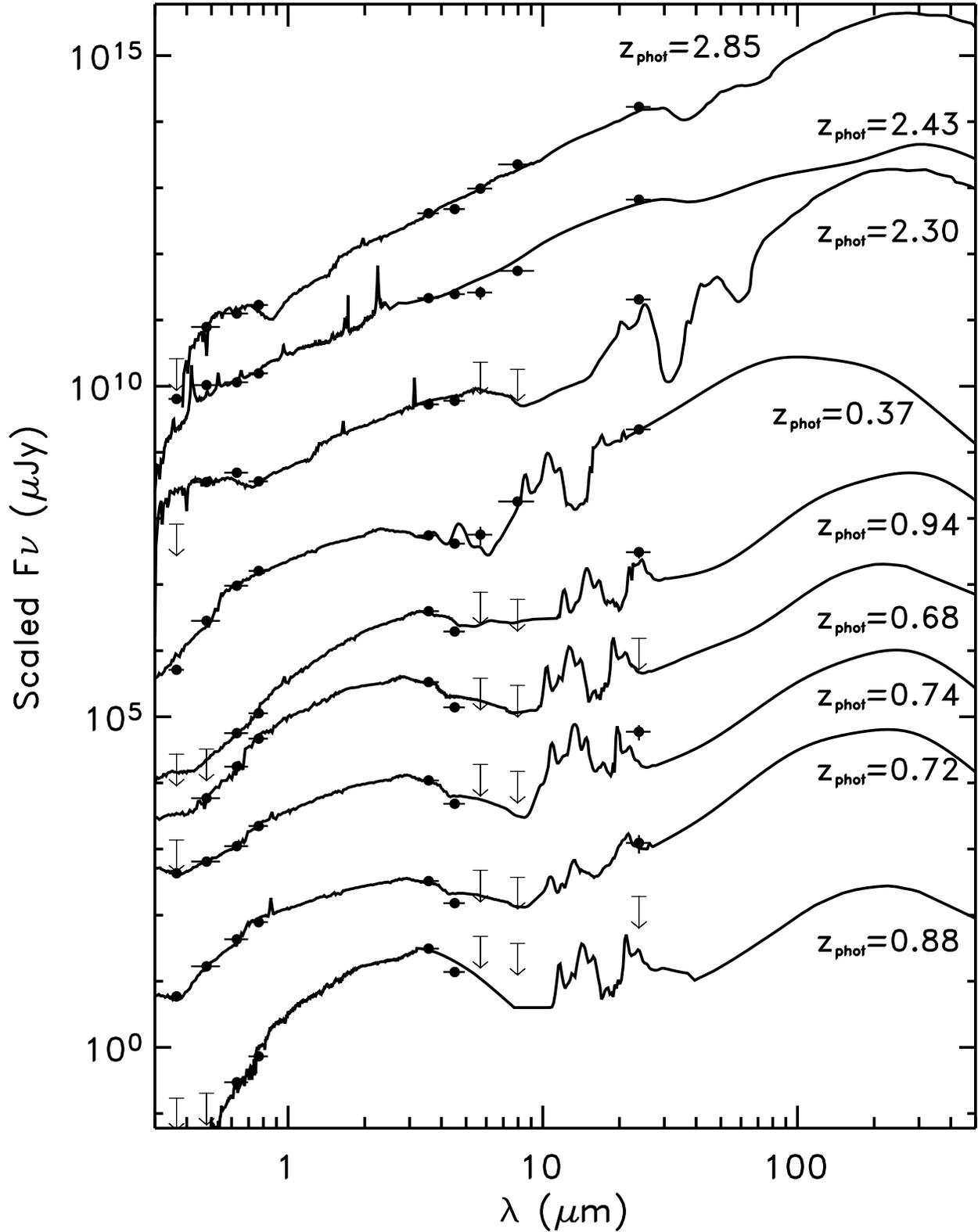}
\caption{SEDs for 5 sources with [3.6$-$4.5]$<-$0.3 (lower 5 SEDs) and 
4 sources redder than [3.6$-$24]=7.5 mag.  In most cases the uncertainties 
are smaller than the symbol sizes.  
}
\end{figure}
\clearpage

We have also selected all sources redder than [3.6$-$24]=7.5 mag. for 
investigation (see Figure 2), requiring a detection at S/N$>$5 at 24$\mu$m. 
Of 63 sources with [3.6$-$24]$>$7.5 mag., 42 were found to have valid colors 
this red on re-derivation of their fluxes by hand.  The remainder are
about evenly divided between sources for which 
more than one 3.6$\mu$m source likely contributes to the larger beam
24$\mu$m emission (a commonly expected situation due
to the large difference in beam profiles), 
and spuriously low 3.6$\mu$m flux densities
caused by latents or electronic offsets due to nearby bright stars.
This latter category of anomalies at 3.6$\mu$m represents a 10/16075=0.06\% 
anomaly rate amongst the whole catalog, and 17\% among the selected red 
sources.  As for the unusually blue [3.6$-$4.5] sources, a high anomaly rate
amongst color outliers is not unanticipated.

\clearpage
\begin{deluxetable}{lcccccclc}
\tabletypesize \footnotesize
\tablewidth{0pt} 
\tablecaption{Selected Sources with Unusually Blue [3.6$-$4.5] or Red [3.6$-$24] Colors}
\tablehead{
\colhead{Name} & \colhead{RA} & \colhead{Dec} & \colhead{z$_{phot}$} & \colhead{L$_{3-1000{\mu}m}$} & 
 \colhead{A$_V$} & \colhead{Template} & \colhead{Deviation, $\sigma$} \\
\colhead{} & \colhead{J2000} & \colhead{} & \colhead{} & \colhead{L$_{\odot}$} & 
 \colhead{mag.} & \colhead{} & \colhead{blue sources} \\
}
\startdata
SWIRE\_J104513.3$+$585933 & 10 45 13.39  & 58 59 33.5  & 0.88 & 10.3 & 0.8 & Sa & 7.2 \\
SWIRE\_J104552.8$+$590600 & 10 45 52.86  & 59 06 00.8  &	0.72 & 11.3 & 1.0 & Sd pec & 7.2 \\
SWIRE\_J104657.3$+$590902 & 10 46 57.38  & 59 09 02.5  &	0.74 & 10.9 & 0.2 & Sdm & 4.6 \\
SWIRE\_J104743.7$+$591034 & 10 47 43.75  & 59 10 34.6  &	0.68 & 10.8 & 0.5 & Sbc HII & 8.4 \\
SWIRE\_J104436.8$+$591349 & 10 44 36.84  & 59 13 49.2  &	0.94 & 11.2 & 1.3 & Sc strbst & 6.2 \\
SWIRE\_J104616.0$+$591424 & 10 46 16.08  & 59 14 24.9  &	0.37 & 11.1 & 0.8 & Im pec HII & \nodata \\
SWIRE\_J104511.8$+$590121 & 10 45 11.88  & 59 01 21.6  &	2.30 & 13.4 & 0.2 & HII  & \nodata \\
SWIRE\_J104613.4$+$585941 & 10 46 13.44  & 58 59 41.3  &	2.43 & 13.2 & 1.1 & QSO  & \nodata \\
SWIRE\_J104700.2$+$590107 & 10 47 00.20  & 59 01 07.6  &	2.85 & 13.7 & 0.3 & Sy1  & \nodata \\
\enddata
\end{deluxetable}
\clearpage

The best fit redshifts for 4 representative red sources (Figure 3 upper 4
SEDs \& Table 2 last 4 entries) range from 0.37 to 2.85, with a luminosity
range of log L$_{3-1000{\mu}m}$=11.1 to 13.7 L$_{\odot}$.  It is very
difficult to obtain unique fits for some objects of this type owing to the
flatness of the SEDs and the limited number of data points, and
these fits should be regarded as illustrative only, pending a thorough
analysis of the possible range of templates, redshifts and luminosities
that can fit each of these sources.  These objects appear to be starbursts,
ULIRGs and AGN with a wider redshift and luminosity range than the blue
sources in Figure 3, including some z$>$2 objects with luminosities in the
hyperluminous object (HyLIRG) range.  SWIRE is expected to be particularly
sensitive to high-redshift IR-luminous AGN, which are expected to be bright
in the very sensitive 24$\mu$m band due to warm circumnuclear dust.  The
high-redshift volume density of HyLIRGs will be important for
models for the early formation of massive systems in the Universe.  Spitzer
IRS spectroscopy may prove essential for determining redshifts and
excitations for the reddest, optically-faintest, systems.

\clearpage
\begin{figure}
\plotone{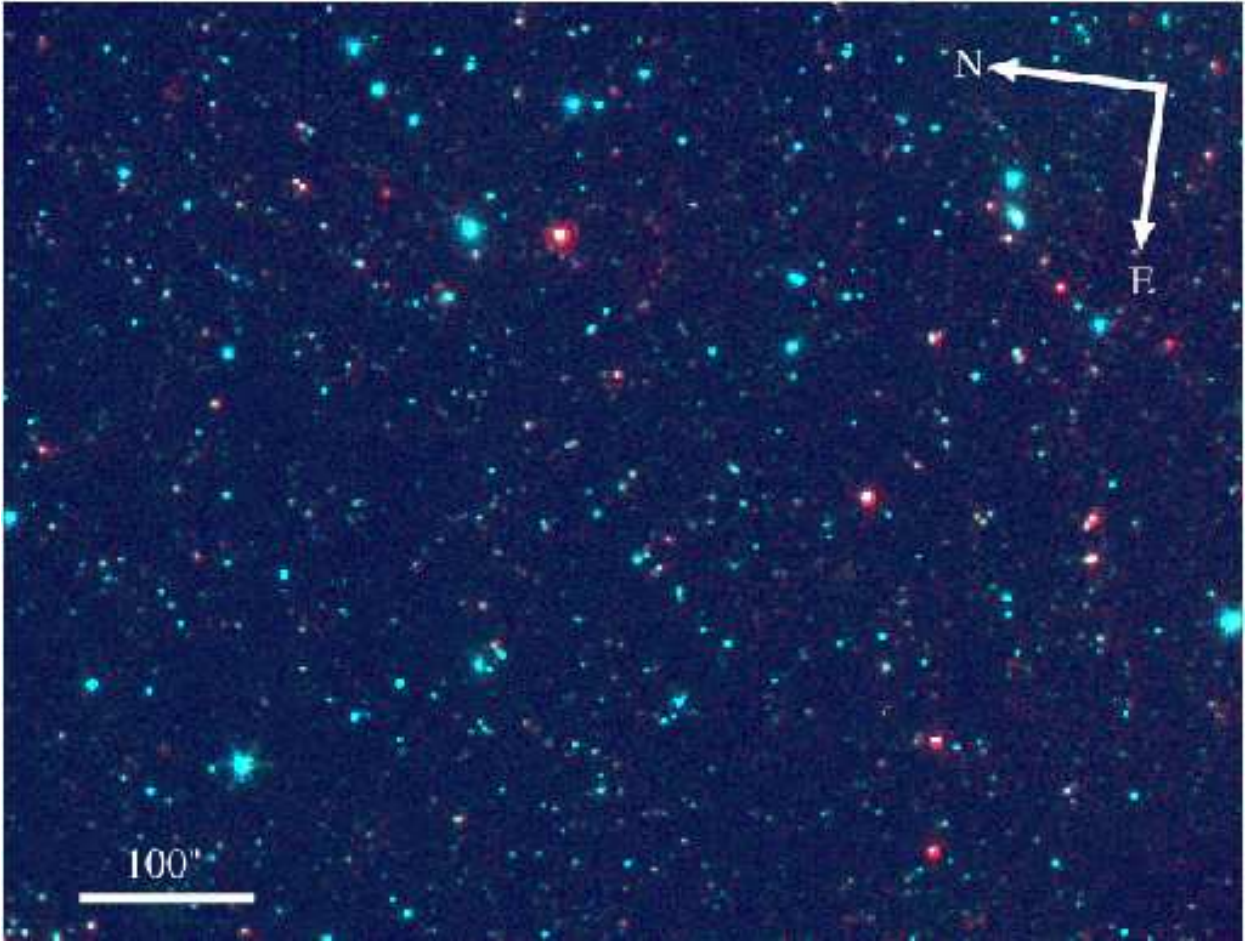}
\caption{Plate 1: 3-color image of $\sim$0.03 sq. deg of the SWIRE Lockman 
validation 
field, centered at $10^h47^m32.67^s$,  $59^d07^m16.3^s$: 3.6$\mu$m (blue),
4.5$\mu$m (green), 24$\mu$m (red).
}
\end{figure}
\clearpage
\acknowledgments
Support for this work, part of the Spitzer Space Telescope Legacy Science 
Program, was provided by NASA through an award issued by the Jet Propulsion 
Laboratory, California Institute of Technology under NASA contract 1407.

\end{document}